\documentclass[twocolumn,twoside,slac_two]{revtex4}
\usepackage{graphicx}
\usepackage{fancyhdr}
\usepackage{psfrag}

\newcommand{\nl}{\nonumber\\}
\newcommand{\order}{{\cal O}}

\newcommand{\be}{\begin{equation}}
\newcommand{\ee}{\end{equation}}

\newcommand{\bea}{\begin{eqnarray}}
\newcommand{\eea}{\end{eqnarray}}

\begin{document}

\title{The Modern Description of Semileptonic Meson Form Factors}

\author{Richard J. Hill}
\affiliation{
Fermi National Accelerator Laboratory \\
P.O. Box 500, Batavia, IL 60510, U.S.A.
}

\begin{abstract}
I describe recent advances in our understanding of the hadronic form 
factors governing semileptonic meson transitions.  The resulting framework
provides a systematic approach to the experimental data, as a means of 
extracting precision observables, testing nonperturbative field theory
methods, and probing a poorly understood limit of QCD. 
\end{abstract}

\maketitle

\thispagestyle{fancy}

\section{Introduction: into the meson\label{sec:intro}}

Semileptonic transitions of one meson into another 
yield important measurements of both weak and strong dynamics. 
By comparing the experimentally determined 
decay rate to a theoretical normalization of the relevant 
hadron transition
amplitude at one or more kinematic points, elements of the 
Cabibbo-Kobayashi-Maskawa (CKM) matrix are determined.   
Independent of the overall normalization, the shape of the 
semileptonic spectrum provides a quantitative probe of
underlying hadron dynamics. 

Of the six CKM elements that can be probed directly using stable hadrons,
determinations from exclusive semileptonic transitions 
are the most precise ($|V_{us}|$); 
competitive with other determinations 
($|V_{ub}|$, $|V_{cb}|$ from {\it inclusive} semileptonic $B$ decays;  
$|V_{cd}|$ from deep-inelastic neutrino scattering; 
$|V_{cs}|$ from charm-tagged $W$ decays); 
or complementary to existing determinations 
($|V_{ud}|$, from nuclear beta decay)~\cite{Eidelman:2004wy}. 
In all cases, the theoretical normalization
gives a dominant error.  The experimentally determined spectrum can 
be used both to test the nonperturbative methods used in determining
this normalization, and to optimize the merger of theory with experiment. 

\begin{figure}[ht]
\centering
\includegraphics[width=60mm]{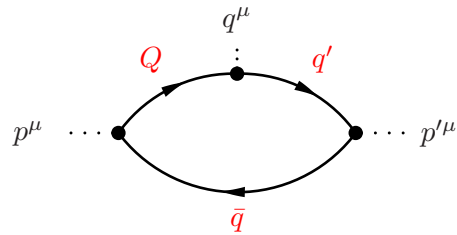}
\caption{Correlation function from which the semileptonic form factors 
are extracted.  $Q$, $q^\prime$ and $q$ are quark labels. $p^\mu$, $p^{\prime\mu}$ 
and $q^\mu$ are momenta.  
} \label{fig:onebody}
\end{figure}

For the study of hadron dynamics, the use of a virtual $W$ boson 
probe in semileptonic transitions can be viewed 
in analogy with the use of a virtual photon in deep-inelastic scattering. 
The ``known'' and theoretically ``clean'' weak (electromagnetic) physics  
is used to probe the ``unknown'' and theoretically ``messy'' strong physics of the 
meson (proton).  
For the case of semileptonic transitions, 
in addition to the virtuality of the exchanged boson, we can envision 
``dialing the knobs'' of the initial- and final-state meson 
masses.
To be precise, consider the situation pictured in Fig.~\ref{fig:onebody}. 
A $Q \bar{q}$ pair with quantum numbers of the ``heavy'' meson $H$ is created at spacetime point
$x$, interacts with a flavor-changing current at $y$, and the 
$q^\prime \bar{q}$ pair with quantum numbers of the ``light'' meson $L$ is annihilated at $z$. 
By suitable manipulations~%
\footnote{ 
Fourier transform, $x\to p$, $z\to p^\prime$, and
tune $p^2\to m_H^2$, $p^{\prime 2} \to m_L^2$
}, 
we can extract hadronic transition amplitudes (form factors) from
this correlation function.   
The entire process takes place in the complicated QCD background with soft and 
hard gluons interacting with everything, $q\bar{q}$ pairs popping out of the vacuum, 
etc.  On this fixed background, we can compute a correlation function with  
any quark masses we desire; in fact Nature has chosen a few fixed values (and 
presumably obtained the correct answer), and we shall be content with these.~%
\footnote{
(Unquenched) lattice simulations study essentially the same object.  However, in contrast to
the fixed background described here, the ``valence'' quark masses injected by the
currents are coupled to the dynamical ``sea'' quark masses.  When extrapolated to
the physical masses, the results must of course be the same. 
}

In this way, a rather complete exploration of the ``1-body'' semileptonic topology, 
summarized by invariant form factors $F(m_H,m_L,q^2)$, is possible. 
A similar analysis can be applied to ``0-body'' leptonic decays, 
and to ``2-body'' hadronic decays.  
The former case is rather simple: by kinematics there is only the single ``knob'' 
of the meson mass, and the result is summarized by a single number, the decay constant 
$f(m_H)$.  The latter case is rather complicated, with many different topologies 
contributing to a typical physical process~\cite{Buras:1998ra}.  

The remainder of the talk is organized as follows.  
Section~\ref{sec:analyticity} reviews the first-principles knowledge we
have about the form factors, following just from kinematics without dynamics. 
Pseudoscalar-pseudoscalar transitions between ``heavy-light'', nonsinglet mesons
are particularly simple and are the main focus.~%
\footnote{
The nonsinglet restriction ensures that only a single topology is relevant as 
in Figure~\ref{fig:onebody}. 
}   
Rigorous power-counting arguments provide the basis for a powerful expansion 
based on analyticity. 
Section~\ref{sec:data} illustrates how the experimental data is simplified
by making use of this expansion.  
In particular, we find the remarkable conclusion that in terms of standard 
variables, no semileptonic meson form factor has ever been observed 
to deviate from a straight line. 
Given that the form factors are indistinguishable from straight lines, if
the shape of the semileptonic spectrum is to provide insight on QCD, 
it must be through the slope of the form factor; in fact, a clear but unsolved
question in QCD translates directly into the numerical value of this slope in an 
appropriate limit, as described in Section~\ref{sec:fundamental}.  
Phenomenological implications in the $B\to\pi$ system are considered in 
Section~\ref{sec:pheno}. 
The methodology described here provides a convenient framework in which to understand 
precisely what measurements in the charm system can, and cannot, say that is relevant to
the bottom system, as discussed in Section~\ref{sec:charm}. 
Section~\ref{sec:pv} outlines the extension to 
pseudoscalar-vector transitions. 

\section{Analyticity and crossing symmetry \label{sec:analyticity} }

An oft-cited downside of old and well-known dispersion-relation arguments 
is that the results are too general, and do not make specific 
predictions for detailed dynamics.   
In fact, precisely these properties make them useful to the 
problem at hand---it is essential to make {\it some} statement on the possible 
functional form of the form factors, yet we do not want to make assumptions, 
explicit or implicit, on the dynamics. 

\begin{table}[t]
\begin{center}
\caption{Maximum $|z(t,t_0)|$ throughout semileptonic range 
with symmetrizing choice $t_0=t_+(1-\sqrt{1-t_-/t_+})$.}
\begin{tabular}{ccc}
\textbf{\quad Process \quad}  & \textbf{CKM element} & \textbf{$|z|_{\rm max}$}
\\
\hline 
$\pi^+ \to \pi^{0}$ & $V_{ud}$ & $3.5 \times 10^{-5}$ \\
$B \to D$ & $V_{cb}$ & 0.032 \\
$K \to \pi$ & $V_{us}$ & 0.047 \\
$D \to K$ & $V_{cs}$ & 0.051 \\
$D \to \pi$ & $V_{cd}$ & 0.17 \\
$B \to \pi$ & $V_{ub}$ & 0.28 
\end{tabular}
\label{tab:zmax}
\end{center}
\end{table}

\begin{figure}[h]
\centering
\includegraphics[width=80mm]{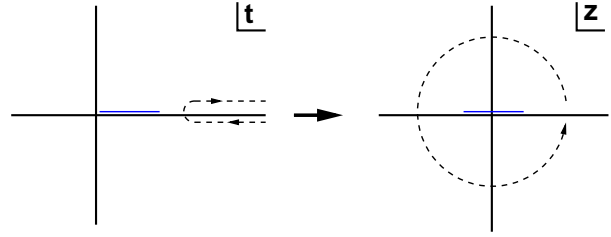}
\caption{Mapping (\ref{eq:zdef}) 
of the cut $t$ plane onto the unit circle.  The semileptonic 
region is represented by the blue line.} \label{fig:cutplane}
\end{figure}

The analytic structure of the form factors can be investigated by 
standard means.~%
\footnote{ 
For a general discussion, see e.g. \cite{bjorken}.  For early work on 
applications to 
semileptonic form factors, see \cite{Bourrely:1980gp,Boyd:1994tt,Lellouch:1995yv,Boyd:1995sq,Caprini:1995wq,Caprini:1997mu,Boyd:1997qw}. 
}
Let us focus on the form factors for 
pseudoscalar-pseudoscalar transitions, defined by the matrix element of the
relevant weak vector current, ($q\equiv p-p^\prime$) 
\begin{eqnarray}\label{eq:vectorff}
&& \langle L(p^\prime)| V^\mu | H(p)\rangle  \nl
&&\quad = F_+(q^2)\left( p^\mu + p^{\prime\mu}\right) + F_-(q^2)q^\mu  \nl 
&&\quad = F_+(q^2)\left( p^\mu + p^{\prime\mu} - {m_H^2-m_L^2\over q^2}q^\mu \right)  \nl 
&& \qquad\qquad   
+ F_0(q^2){m_H^2-m_L^2\over q^2}q^\mu \,. 
\end{eqnarray}
To ensure that there is no singularity at $q^2=0$, the form factors obey
the constraint 
\be\label{eq:kinconstraint}
F_+(0) = F_0(0) \,. 
\ee

Ignoring possible complications from 
anomalous thresholds or subthreshold resonances, 
to be discussed below, the form factors $F(t=q^2)$ 
can be extended to analytic functions throughout the complex $t$ plane, except 
for a branch cut along the positive real axis, starting at the point $t=t_+$ 
[$t_\pm \equiv (m_H\pm m_L)^2$]
corresponding to the threshold for production of real $\bar{H}L$ pairs in 
the crossed channel. 
By a standard transformation, 
as illustrated in Figure~\ref{fig:cutplane}, 
the cut $t$ plane is mapped 
onto the unit circle $|z|\le 1$, 
\begin{equation}\label{eq:zdef}
z(t,t_0) \equiv { \sqrt{t_+-t} - \sqrt{t_+-t_0} \over \sqrt{t_+-t} + \sqrt{t_+-t_0}} \,,
\end{equation} 
where $t_0$ is the point mapping onto $z=0$.  
The isolation of the semileptonic region from singularities in the $t$ plane 
implies that $|z|<1$ throughout this region.  
Choosing $t_0=t_+(1-\sqrt{1-t_-/t_+})$ minimizes the maximum value of $|z|$; for 
typical decays these maximum values are given in Table~\ref{tab:zmax}.

Since the form factor is analytic, it may be expanded, 
\begin{equation}\label{eq:Fexpand}
F(t) = {1\over P(t)\phi(t,t_0)} \sum_{k=0}^\infty a_k(t_0) z(t,t_0)^k \,,
\end{equation} 
where $P$ and $\phi$ will be explained shortly.  
From Table~\ref{tab:zmax},  it is apparent that if 
some control over the coefficients $a_k$ can be established, 
the expansion is rapidly convergent.  

To bound the coefficients appearing in (\ref{eq:Fexpand}), 
we consider the norm, 
\begin{eqnarray}\label{eq:norm}
|| F ||^2 &\equiv& \sum_{k=0}^\infty a_k^2  = {1\over 2\pi i} \oint {dz\over z} |P\phi F|^2 \nl
&=& {1\over \pi}\int_{t_+}^\infty {dt \over t-t_0} \sqrt{ t_+-t_0 \over t - t_+} |P\phi F|^2 \,.
\end{eqnarray}
By crossing symmetry, the norm can be evaluated using the form factors 
for the related process of $\bar{H}L$ production.    

\subsection{Subthreshold poles, (absence of) anomalous thresholds, and a choice of $P$} 

For some hadronic processes, it may happen that subthreshold resonances occur in the
production amplitude, which must be properly taken into account.  
Particles lying below threshold are hadronically stable, so that
ignoring higher-order weak and electromagnetic corrections, they are described 
by simple poles.  
The canonical example is the $B^*$ pole appearing in the vector channel for $B\to\pi$. 
Such poles could 
in principle be simply subtracted, but doing so requires knowledge of the relevant 
coupling appearing as the coefficient of $1/(m_{B^*}^2 -t)$ in the dispersive 
representation of the form factor.   
Armed with only the knowledge of the pole position, 
this pole can instead be removed by multiplying with a function $P(t)$ with a simple
zero at $m_{B^*}^2$.  Requiring also that the function satisfy $|P|=1$ along the cut, 
up to an arbitrary phase, 
\begin{equation}\label{eq:pdef}
P_{F_+}^{B\to\pi}(t) = z(t,m_{B^*}^2) \,,
\end{equation}
with $z$ as in (\ref{eq:zdef}). 

It may happen in some processes that ``anomalous thresholds'' appear. 
The relevant aspects of this technical subject can be summarized as follows:  
an anomalous threshold 
can occur in the spacelike region, $t<0$, only if $H$ and $L$ are unstable, i.e., 
$m_{H\,{\rm or}\,L} > m_X + m_Y$ for some hadrons $X$ and $Y$ coupling to $H$ or $L$. 
Similarly, an anomalous threshold can occur in the timelike region, 
$t_-< t <t_+$,
only if $m_{H \,{\rm or}\, L}^2 > m_X^2 + m_Y^2$.~%
\footnote{
In geometrical language, this can be related to the statement that 
a triangle cannot have more than one obtuse angle~\cite{bjorken}. 
Strictly speaking, for ``heavy-to-heavy'' transitions such as $D\to K$ and 
$B\to D$, additional Zweig-suppressed topologies can also lead to anomalous thresholds, 
related to processes such as $D_s^{*+} \to D_s^{+} \pi^0$.  As indicated by the small
branching fraction, such effects are highly suppressed; for further discussion and 
references, see Section~\ref{sec:pv}.    
}
This explains the priveleged position of ground-state heavy-light pseudoscalar 
mesons.  
The ground state meson with given flavor quantum numbers is necessarily 
pseudoscalar~\cite{Weingarten:1983uj}. 
The mesons $H$ and $L$ are therefore the lightest hadrons containing their 
respective ``heavy'' quarks, so that in particular  
$m_{H\,{\rm or}\,L} < m_X + m_Y$ and 
$m_{H \,{\rm or}\, L}^2 < m_X^2 + m_Y^2$ for any hadrons $X$ and $Y$ containing
the same heavy quark.~%
\footnote{  
The same is not true for ``heavy-heavy'' systems; e.g. a $Q\bar{Q}$ pair 
has mass $\sim (2m_Q)^2 \sim 4m_Q^2$, compared to $m_Q^2+m_Q^2 \sim 2m_Q^2$ for 
a pair of ($Q\bar{q}$) mesons.
}
(For the present purposes, this can be taken as the definition of a ``heavy-light''
meson.)
Fortunately, and for related reasons, these mesons are easily 
produced and studied experimentally, and in lattice simulations.

\subsection{Unitarity and a choice of $\phi$} 

Nothing in (\ref{eq:Fexpand}) or (\ref{eq:norm}) yet 
singles out a choice of $\phi$; indeed any analytic function will work, 
(of which there are many!).  
A default choice is determined from arguments based on unitarity: 
by an appropriate choice of $\phi$, 
the norm can be identified as a partial rate for some inclusive 
process that is perturbatively calculable.   In particular, from 
\begin{eqnarray}\label{eq:ope}
\!\!\Pi^{\mu\nu}(q) &\equiv& i \int d^4x\, e^{iq\cdot x} \langle 0 | T\left\{ V^\mu(x) V^{\nu\dagger}(0) \right\} | 0 \rangle  \nl
&=& (q^\mu q^\nu - g^{\mu\nu} q^2)\Pi_1(q^2) + q^\mu q^\nu \Pi_0(q^2) \,, 
\end{eqnarray}
unsubtracted dispersion relations can be written for the quantities ($Q^2= -q^2$) 
\begin{eqnarray}\label{eq:subtracted}
\chi_{F_+}(Q^2) &=& {1\over 2}{\partial^2 \over \partial (q^2)^2} \left[ q^2 \Pi_1 \right] 
= {1\over \pi} \int_0^\infty\! dt\, {t {\rm Im}\Pi_1(t) \over (t+Q^2)^3 } \,, \nl
\chi_{F_0}(Q^2) &=& {\partial \over\partial q^2} \left[ q^2 \Pi_0 \right] 
= {1\over \pi}  \int_0^\infty\! dt\, {t {\rm Im}\Pi_0(t) \over (t+Q^2)^2 } \,. \nl
\end{eqnarray} 
Noticing that for $t>t_+$, ($\eta$ an isospin factor)
\begin{eqnarray}\label{eq:inequality}
{\eta \over 48\pi}{ [(t-t_+)(t-t_-)]^{3/2} \over t^3} |F_+(t)|^2
&\le& {\rm Im} \Pi_1(t) \,, \nl
{\eta t_+ t_- \over 16\pi}{ [(t-t_+)(t-t_-)]^{1/2} \over t^3} |F_0(t)|^2 &\le& 
{\rm Im} \Pi_0(t) \,, 
\end{eqnarray} 
shows that an upper bound on the norm can be established 
by choosing [recall that $|z|=1$ along the integration contour in (\ref{eq:subtracted})]
\begin{eqnarray}\label{eq:phi}
\phi_{F_+}(t,t_0) &=& \sqrt{\eta\over 48\pi}  
{t_+ - t \over (t_+ - t_0)^{1/4} } 
\left( z(t,0)\over -t \right) \nl
&&  \times
\left( z(t,-Q^2) \over -Q^2 - t \right)^{3/2}
\left( z(t,t_0) \over t_0 - t \right)^{-1/2} \nl
&& \times 
\left( z(t,t_-) \over t_- - t \right)^{-3/4} \,, 
\nl
\phi_{F_0}(t,t_0) &=& \sqrt{\eta t_+ t_- \over 16\pi} 
{\sqrt{t_+ - t} \over (t_+ - t_0)^{1/4} } 
\left( z(t,0) \over -t \right) \nl 
&& \times 
\left( z(t,-Q^2) \over -Q^2 - t \right)
\left( z(t,t_0) \over t_0 - t \right)^{-1/2} \nl
&& \times
\left( z(t,t_-) \over t_- - t \right)^{-1/4}  \,. 
\end{eqnarray} 

The choice of subtracted dispersion relation in 
(\ref{eq:subtracted}) leads to a ``default'' choice
for $\phi$ in (\ref{eq:phi}).  
With this choice, power counting shows that 
$a_k/a_0$ and $\sum_k (a_k^2/a_0^2)$ do not scale as powers 
of large ratios such as $Q/\Lambda_{\rm QCD}$, or $m_Q/\Lambda_{\rm QCD}$ 
when a heavy-quark mass is present~\cite{Becher:2005bg}.  
This ensures that there is no parametric enhancement of the 
coefficients $a_k$ that could offset the smallness of $z^k$ 
in the series (\ref{eq:Fexpand}).   
In fact, at sufficiently large $k$ the coefficients must {\it decrease}  
in order that the sum of squares converge. 
These properties [analyticity, and $a_k/a_0 \sim \sum_k a_k^2/a_0^2 \sim \order(1)$ ] 
are all that is required from the choice of 
$\phi$.  
The ``physical'' prescription following from (\ref{eq:subtracted}),
(\ref{eq:inequality}) and (\ref{eq:phi})
automatically ensures that this is the case.   

The original motivation for considering the operator product expansion
(OPE) in (\ref{eq:ope}) is to place a restriction on the coefficients 
in (\ref{eq:norm}) 
according to ($\chi = \chi_{F_{+,0}}$ as appropriate)
\begin{equation} \label{eq:unitarity} 
\sum_{k=0}^\infty a_k^2 \le \chi(Q^2) \,. 
\end{equation} 
However, in order that an OPE expansion for $\chi(Q^2)$ 
converge, $Q^2$ (or $m_Q^2+Q^2$ when a heavy-quark is present) 
must necessarily be large compared to $\Lambda_{\rm QCD}$.  
This results in a bound that is typically overestimated
by some power of the large ratio of perturbative to hadronic scales.
In practice, the numerical value for the 
bound (\ref{eq:unitarity}) itself is largely irrelevant.   What is 
important is that the choice of $\phi$ which it motivates has the
desired properties. 

Having chosen a ``default'' $\phi$, 
for definiteness, we will also take $Q=0$ in (\ref{eq:phi}) 
as the ``default'' choice, and where a particular choice 
is necessary, $t_0 = t_+(1-\sqrt{1-t_-/t_+})$.  

\section{What the data say\label{sec:data}}

\begin{figure}[h]
\centering
\includegraphics[width=80mm]{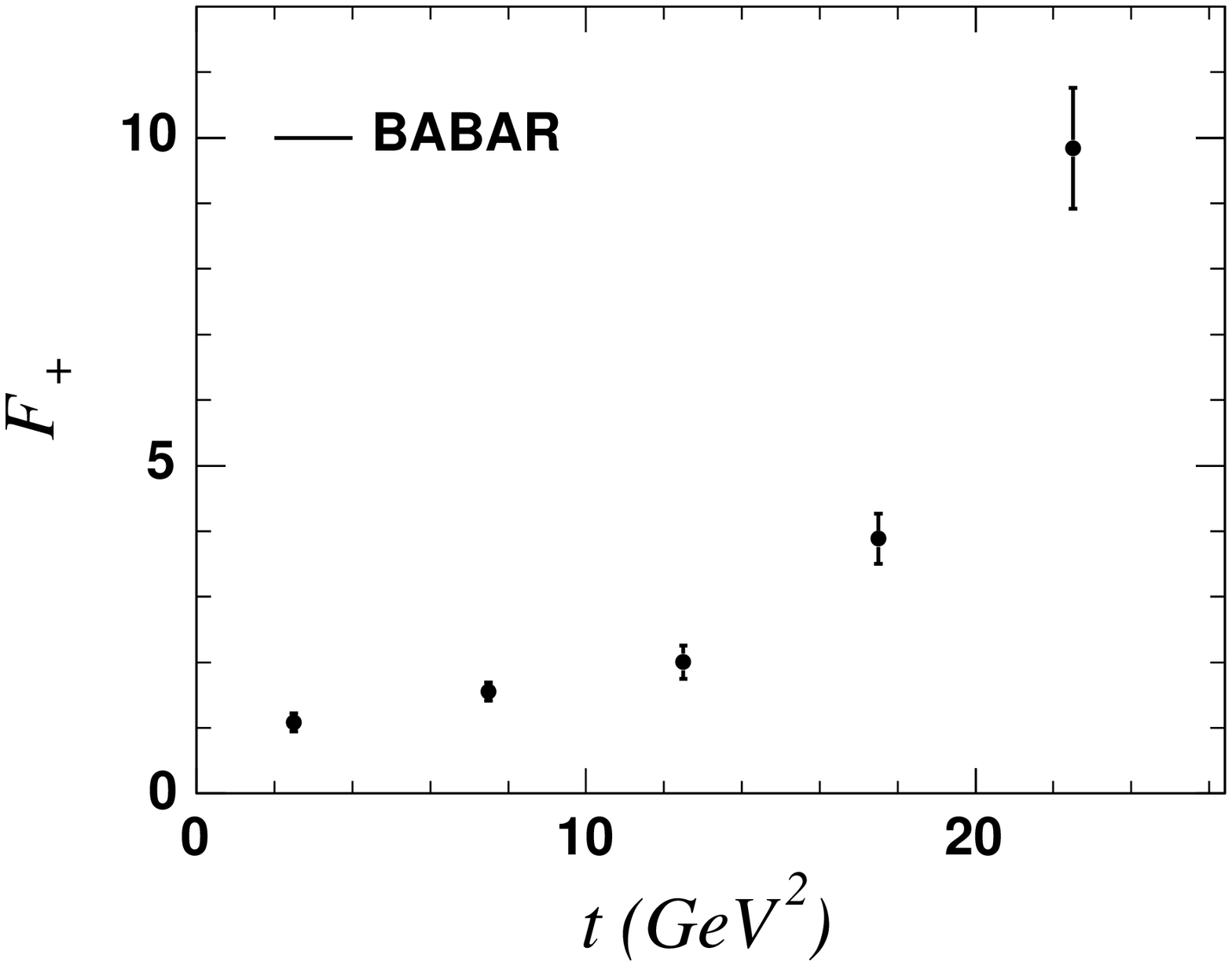}
\includegraphics[width=80mm]{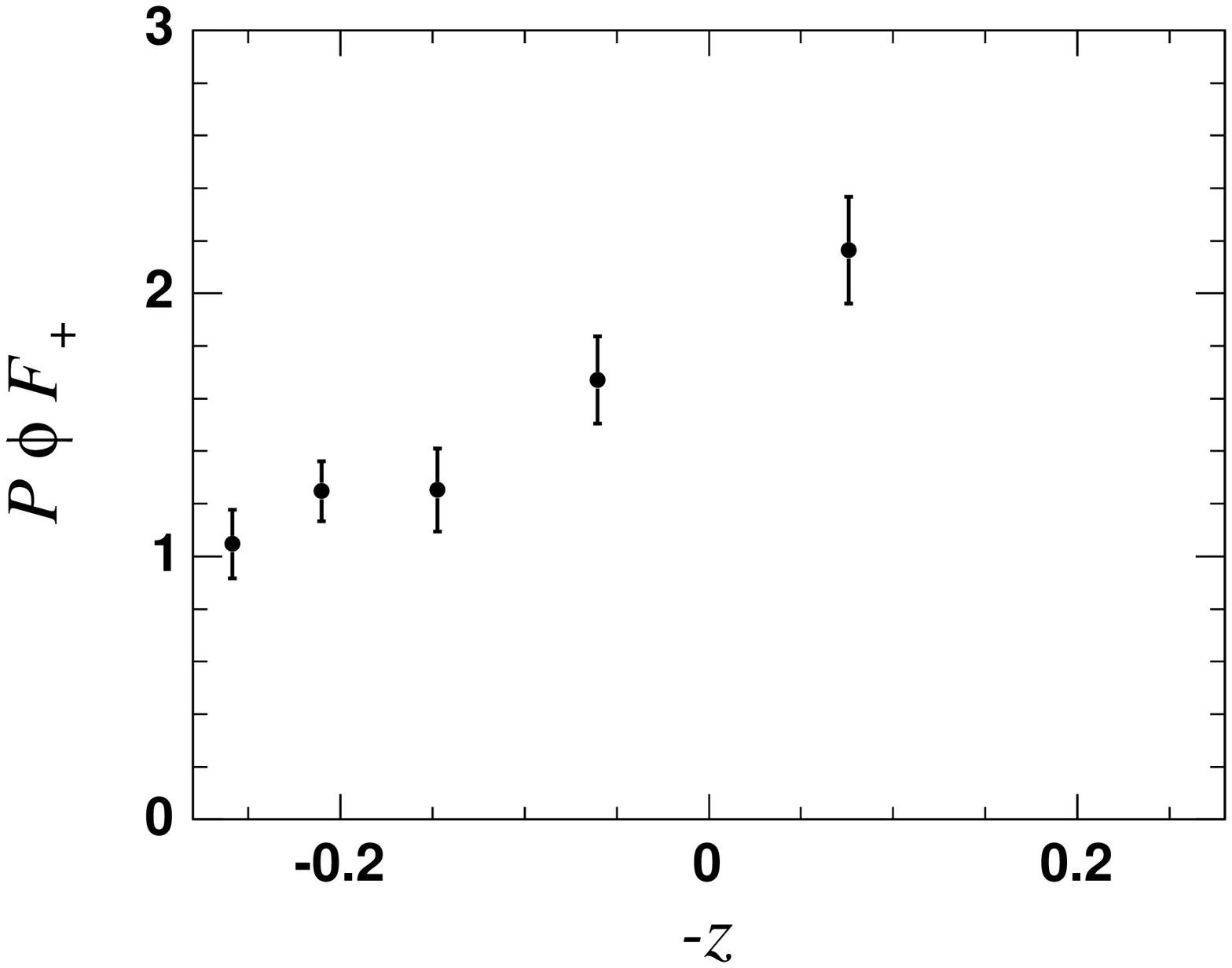}
\caption{Experimental data for $F_+$ in $B\to\pi\ell\nu$, as 
a function of $t$, and for the function $P\phi F_+$ as a function of $z$.    
Partial branching fractions have been converted 
to values of the form factor at the midpoint of each bin in $t$ from
\cite{Aubert:2005cd}.  Units on the vertical axis are arbitrary 
(approximately normalized to unity at $t=0$).  
 } \label{fig:plotqz}
\end{figure}

\begin{table}[t]
\begin{center}
\caption{Linear expansion coefficient $a_1/a_0$ from (\ref{eq:Fexpand}) 
at $t_0=t_+(1-\sqrt{1-t_-/t_+})$ and $Q=0$.  
  }
\begin{tabular}{clc}
\textbf{\quad Process \quad}  & \textbf{$a_1/a_0$} & \textbf{Reference} 
\\
\hline 
%
%
$B \to D$ & $-2.6 \pm 2.3$ & \cite{Abe:2001yf} \\
$K^{+} \to \pi^{0}$ & $-0.2 \pm 0.2$ & \cite{Yushchenko:2004zs} \\
$K_L \to \pi^{\pm}$ & $-0.5 \pm 0.2$ & \cite{Alexopoulos:2004sy} \\
& $\phantom{-}0.0 \pm 0.3$ & \cite{Lai:2004kb} \\
& $-0.2 \pm 0.2$ & \cite{Ambrosino:2006gn} \\
$D \to K$  & $-2.7 \pm 0.5 \pm 0.4$ & \cite{Huang:2004fr} \\
 & $-2.2 \pm 0.4 \pm 0.4$ & \cite{Link:2004dh}  \\
& $-3.2 \pm 0.5 \pm 0.2$ & \cite{Abe:2005sh} \\
$D \to \pi$ & $-2.3 \pm 0.7 \pm 1.3$ & \cite{Huang:2004fr} \\
& $-1.6 \pm 0.5 \pm 1.0$ & \cite{Abe:2005sh} \\
$B \to \pi$ & $-1.3 \pm 0.6 \pm 2.3$ & \cite{Athar:2003yg} \\
& $-1.9 \pm 0.3 \pm 1.1$ & \cite{Aubert:2005cd} \\
& $-1.3 \pm 0.8 \pm 2.2$ & \cite{Abe:2004zm} 
\end{tabular}
\label{tab:a1}
\end{center}
\end{table}

Table~\ref{tab:zmax} can be used to predict the level of 
precision at which slope, curvature, and higher-order corrections 
can be resolved by the data.  
With the ``default'' values of $Q$ and $t_0$, Table~\ref{tab:a1}
shows the results for $a_1/a_0$ obtained from data.  
Except where indicated, modes related by isospin are combined. 
For the $K\to\pi$ case, the results of 
\cite{Yushchenko:2004zs,Alexopoulos:2004sy,Lai:2004kb,Ambrosino:2006gn} 
were presented as a simple quadratic Taylor expansion of the form factor 
about $t=0$.~%
\footnote{
It is desirable to fit the data directly to (\ref{eq:Fexpand}), to avoid biases 
introduced by the truncated $t$ series~\cite{hill}.  
}  
These results have been
converted to the quadratic $z$ parameterization in (\ref{eq:Fexpand}), 
by identifying the Taylor series at $t=0$, 
and propagating errors linearly.   
For $B\to D$, the results of \cite{Abe:2001yf} were presented in terms of
a parameterization 
obtained by expanding $\phi$ and $P$ as a Taylor series in $z$ \cite{Caprini:1997mu}. 
The result in Table~\ref{tab:a1} is obtained by converting  
to the linear $z$ parameterization in (\ref{eq:Fexpand}), with three 
subthreshold ``$B_c^*$'' poles located at 
$m=6.337, 6.899, 7.012\,{\rm GeV}$~\cite{Caprini:1997mu,Eichten:1994gt}, 
and then identifying the coefficients in a Taylor series at $t=t_+(1-\sqrt{1-t_-/t_+})$.  
The results for $D\to K$, $D\to \pi$ and $B\to \pi$ were obtained by fitting 
the linear $z$ parameterization in (\ref{eq:Fexpand}) to the data.  
A second 
error is included by redoing the fits with the quadratic $z$ parameterization, 
subject to the conservative bound $|a_k/a_0|< 10$. 
Due to the smallness of $z$ for pion beta decay, 
$\pi^+\to \pi^0$ (cf. Table~\ref{tab:zmax}), 
the slope in this case is orders of magnitude from being measured
experimentally~\cite{Cirigliano:2002ng}. 

At least through linear order, 
there is no evidence of anomalously large coefficients that could
upset the power counting.  
While it would be desirable to push to the next order and examine the
size of $a_2/a_0$, comparison to data establishes the remarkable 
conclusion that form-factor curvature has not yet been seen  
in any semileptonic transition.  
In fact, for many cases, a form factor slope has yet to be measured.  
An example of the transformed 
form factor is illustrated for $B\to\pi$ in Fig.~\ref{fig:plotqz}. 

From the amusing coincidence that $z^{DK}/z^{D\pi} \sim |V_{cd}|/|V_{cs}|$, and
$z^{BD}/z^{B\pi} \sim |V_{ub}|/|V_{cb}|$, it turns out that the higher
statistics of the Cabibbo-allowed modes ($B\to D$, $D\to K$) 
are offset at linear order 
by the smallness of $z$.  It is thus likely that curvature will eventually 
be measured first in the Cabibbo-suppressed modes ($B\to\pi$, $D\to\pi$). 

The results in Table~\ref{tab:a1} are by no means the final word 
on these quantities, but illustrate the main point, that there is no
sign that the $z$ expansion is breaking down. 
It is also easy to see that unitarity has very little impact. 
For example, for $B\to\pi$, the bound on $F_+$ taken from the OPE at $Q=0$ is overestimated
by a factor $\sim (m_b/\Lambda_{\rm QCD})^3$~\cite{Becher:2005bg}.
Taking for definiteness, $F_+(t_0=16\,{\rm GeV})\approx 0.8$, 
the unitarity bound tells us only that
$\sum_k a_k^2/a_0^2 \lesssim 2500$~\cite{Lellouch:1995yv,Fukunaga:2004zz,Arnesen:2005ez}. 
For $B\to D$, at $Q=0$ with the approximate symmetry relation 
$F_+(t_-) \approx (m_B+m_D)/2\sqrt{m_B m_D}$, and including three subthreshold poles as 
in (\ref{eq:pdef}),  
the unitarity bound is overestimated by a factor $\sim (m_b/m_c)^3$ and yields 
$\sum_k a_k^2/a_0^2 \lesssim 9000$~\cite{Lellouch:1995yv,Caprini:1997mu}.    
While these bounds can be improved somewhat by subtracting off subthreshold poles, 
extending isospin $SU(2)$ to $SU(3)$ flavor symmetry, or by lowering $Q^2$, all of these 
modifications introduce their own uncertainties.~%
\footnote{   
For modes such as $B\to D$, the incredible smallness of $z$, and the judicious use of 
heavy-quark symmetry, 
allows even very conservative 
unitarity bounds to guarantee few-percent level accuracy by keeping only the linear term 
in (\ref{eq:Fexpand})~\cite{Boyd:1994tt,Caprini:1995wq}.  
}

\section{A fundamental question\label{sec:fundamental}}

Given that the form factors (after extracting standard kinematic factors, 
and expressing them in terms of the appropriate standard variable) are 
so far indistinguishable from a straight line, it is apparent that 
any insight to be gained from the shape of the form factors, 
whether it be tests of nonperturbative methods, inputs to other processes, 
or more fundamental questions about QCD, must be based in 
first approximation on the slope of the form factor.  
In fact, this quantity does provide a clear test of lattice QCD,
is an important input to hadronic $B$ decays, 
and in an appropriate limit can provide the answer to a longstanding open 
question about the QCD dynamics governing form factors. 

It is convenient to define the physical shape observables in 
terms of the form factor slopes at $t=0$~\cite{Hill:2005ju},
\begin{eqnarray}\label{eq:betadef} 
{1\over \beta} &\equiv& 
{m_H^2-m_L^2 \over F_+(0)} \left. dF_{0} \over dt\right|_{t=0} \,, \nl
\delta &\equiv& 
1 - {m_H^2-m_L^2 \over F_+(0)} 
\left( \left. dF_{+} \over dt\right|_{t=0}  - \left. dF_{0} \over dt\right|_{t=0} \right) \nl
&\equiv& {F_+(0)+F_-(0)\over F_+(0)}
\,.
\end{eqnarray}
The quantities $\beta$ and $\delta$ depend only 
on the masses of the mesons involved.~%
\footnote{  
Recall that we consider mesons with a fixed light spectator quark, 
which for simplicity in the discussion is assumed massless.   The
meson mass is therefore in one-to-one correspondence with the
heavy (non-spectator) quark mass.
} 
Being physical quantities, they are
independent of any renormalization scale or scheme.  
As discussed in the introduction, these quantities take 
definite values for all $m_H$ and $m_L$, values that 
are accessible experimentally at the fixed masses
$m_\pi$, $m_K$, $m_D$ and $m_B$.~%
\footnote{
For some studies of $\delta - 1 \equiv F_-(0)/F_+(0)$, also called 
$\xi(0)$, in the early literature of light-meson form factors, 
see the review \cite{Chounet:1971yy}. 
The positive sign for $\xi(0)$ predicted in a number of models, 
e.g. \cite{Gershtein:1976aq,Khlopov:1978id}, is in disagreement with current data.   
For some early work on heavy-to-light meson form factors, see \cite{Akhoury:1993uw}. 
The prediction $\lim_{m_B\to\infty} \delta(m_\pi,m_B) = 2$ of 
\cite{Akhoury:1993uw} also appears difficult to reconcile with present data, see below. 
}   
 
Due to the kinematic constraint (\ref{eq:kinconstraint}), 
the difference of form factor slopes is particularly simple.    
Firstly, for $m_H=m_L$, 
\begin{equation}
\delta(m_L,m_H) = 1 \,, \quad [ m_H=m_L ] \,. 
\end{equation}
This is the statement of current conservation ($F_-=0$) in 
(\ref{eq:vectorff}).   
There are three further distinct limits that we can consider: 
$m_H,m_L \gg \Lambda_{\rm QCD}$, $m_H,m_L \ll \Lambda_{\rm QCD}$, and 
$m_H \gg \Lambda_{\rm QCD} \gg m_L$.  Each limit provides valuable insight,
and we consider each in turn. 

\subsection{ $m_H,m_L \gg \Lambda_{\rm QCD}$:  HQET \label{sec:hqet} } 

\begin{figure}[h]
\centering
\includegraphics[width=80mm]{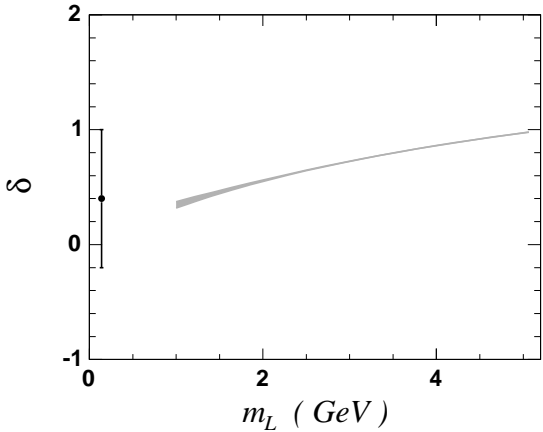}
\caption{ $\delta$ as a function of light-meson mass for fixed heavy-meson mass
$m_H=m_B$.   The shaded band gives the prediction from HQET valid for
$m_H \gtrsim 1\,{\rm GeV}$.  
The dot with error bars is from experimental data as discussed in the text. 
 } \label{fig:hqet}
\end{figure}

For $m_H, m_L \gg \Lambda_{\rm QCD}$, it is convenient to express 
the form factors $F_{\pm}$ 
in terms of reduced amplitudes, $h_{\pm}$ where the dominant heavy-quark 
mass dependence is extracted: [cf. (\ref{eq:vectorff})] 
\begin{eqnarray} 
&&F_+ (p + p^\prime) + F_- (p - p^\prime) \nl 
&& \quad 
\equiv \sqrt{m_H m_L} 
\left[ h_+ (v+v^\prime) + h_- (v-v^\prime) \right] \,,
\end{eqnarray} 
where the meson velocities are $m_H v \equiv p$, $m_L v^\prime \equiv p^\prime$. 
With these definitions, before any approximation, 
\begin{equation}
\delta(m_L,m_H) \equiv {2m_L\over m_H + m_L} { 1 + {h_-\over h_+} 
\over 1 - {m_H-m_L \over m_H+m_L} {h_-\over h_+} } \,. 
\end{equation} 
At leading power, $h_+ \approx \xi$, 
the universal Isgur-Wise function~\cite{Isgur:1989ed}, and $h_- \approx 0$, so that
\begin{eqnarray}\label{eq:deltahqet}
\delta(m_L,m_H) &=& {2m_L\over m_H+m_L}
\big[ 1 + \order(\alpha_s, \Lambda/m_L,\Lambda/m_H) \big] \,, \nl
&& \quad \big[ m_H,m_L \gg \Lambda \big] 
\,. 
\end{eqnarray}
Radiative and power corrections to 
$h_\pm$ can be analyzed systematically with heavy-quark
effective field theory (HQET)~\cite{Neubert:1993mb}.   
From (\ref{eq:deltahqet}), we see that in the regime where 
$m_H,m_L \gg \Lambda_{\rm QCD}$, $\delta$ can take any 
value between zero and unity.~%
\footnote{
Since we are considering quantities at maximum recoil,
the large recoil parameter, 
$(v\cdot v^\prime)_{\rm max} = (m_H^2+m_L^2)/2m_Lm_H$, can upset the 
power counting when $m_H\to\infty$ at fixed $m_L$.  
For arbitrarily small 
$m_L/m_H$, the limiting value $\delta = 2m_L/(m_H+m_L)$ is 
obtained in the limit where $m_L\to\infty$, $m_H\to\infty$, 
$m_L/m_H = {\rm constant}$.  
}
For fixed  $m_H=m_B$, 
Fig.~\ref{fig:hqet} shows the allowed range of $\delta$ in the 
regime of $m_L$ where this expansion is applicable.   
Results for subleading corrections are taken from \cite{Neubert:1993mb}, 
with errors estimated by 
varying the renormalization scale by a factor of two, and assigning
$100\%$ uncertainty to the power corrections. 
The data point for $B\to\pi$ is from (\ref{eq:deltabpi}). 

\subsection{ $m_H,m_L \ll \Lambda_{\rm QCD}$: CHPT \label{sec:chiral} } 

\begin{figure}[h]
\centering
\includegraphics[width=80mm]{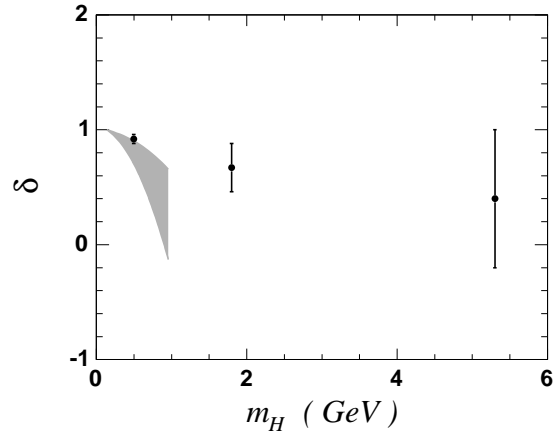}
\caption{ $\delta$ as a function of heavy-meson mass for fixed light meson mass
$m_L=m_\pi$.   The shaded band gives the prediction from CHPT valid for
$m_H \lesssim 1\,{\rm GeV}$.  
Dots with error bars are experimental data as discussed in the text. 
 } \label{fig:chiral}
\end{figure}

For $m_H,m_L \ll \Lambda_{\rm QCD}$, the form factors can be expanded
in powers of the small ratios of masses and momenta relative 
to the QCD scale (since the energies involved in semileptonic
transitions are bounded by meson masses).  
In this regime of Chiral Perturbation Theory 
(CHPT)~\cite{Gasser:1983yg,Gasser:1984gg}, 
$1-\delta$ is of order $m^2/\Lambda_{\rm QCD}^2$.  
The leading contribution is given by 
\begin{eqnarray}\label{eq:deltachiral} 
&&\delta(m_L,m_H) \nl
&& \quad = 1 -  {2(m_H^2-m_L^2)\over F_0^2}
\big[ L_9^r(\mu) - 2 L_5^r(\mu) + \dots \big] \,,  \nl
&& \quad \big[ m_H,m_L \ll \Lambda \big] \,, 
\end{eqnarray} 
where the ellipsis denotes a known kinematic function which 
cancels the renormalization-scale dependence of the low-energy 
constants $L_{5,9}^r(\mu)$.  Here $F_0 \propto f_\pi$ is related to the
pion decay constant.   The deviation of $\delta$ from 
unity is predicted by the sign of the combination $L_9^r - 2L_5^r$, 
which is empirically found to be positive.  
The band in Fig~\ref{fig:chiral} shows the allowed range of $\delta$, 
where for illustration we take~\cite{Gasser:1984gg}
$F_0 = 0.088\,{\rm GeV}$, $10^3 L_5^r(m_\eta) = 2.2 \pm 0.5$ (determined
from the ratio $f_K/f_\pi$) 
 and $10^3 L_9^r(m_\eta) = 7.4 \pm 0.7$ (determined from the electromagnetic
charge radius of the pion).  
The value of $\delta $ for $K^0 \to\pi^+$ is taken from~\cite{Alexopoulos:2004sy}. 
\footnote{
The slope parameters there are related to those in (\ref{eq:betadef}) by 
$1+1/\beta-\delta \approx \lambda_+ (m_K^2-m_\pi^2)/m_\pi^2$ , 
$1/\beta \approx \lambda_0 (m_K^2-m_\pi^2)/m_\pi^2$. 
}
The data point for $D\to\pi$ is discussed in Section~\ref{sec:charm}, and
the data point for $B\to\pi$ is the same as in Fig.~\ref{fig:hqet}. 

\subsection{ $m_H \gg \Lambda \gtrsim m_L$: SCET \label{sec:scet} } 

\begin{figure}[h]
\centering
\includegraphics[width=80mm]{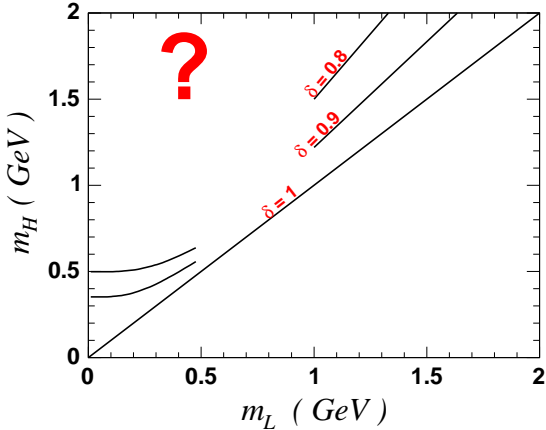}
\caption{Contours of constant $\delta$ in the $m_L - m_H$ plane.   
} \label{fig:contour}
\end{figure}

The HQET description in Section~\ref{sec:hqet} breaks down when the 
light meson becomes light (Fig.~\ref{fig:hqet}).  Similarly, 
the CHPT in Section~\ref{sec:chiral} breaks down when the heavy 
meson becomes heavy (Fig.~\ref{fig:chiral}).   

In this regime, since we are concerned with the point at maximum recoil, 
the light-meson energy necessarily satisfies $E \sim m_H/2 \gg \Lambda_{\rm QCD}$. 
Observables can thus be analyzed using a simultaneous expansion in 
$\Lambda/m_H$ and $\Lambda/E$.  The soft-collinear effective theory (SCET)
framework has been developed to study this 
regime~\cite{Bauer:2000ew,Bauer:2000yr,
Chay:2002vy,Beneke:2002ph,Hill:2002vw}.    
The leading description of the form factors for pseudoscalar-pseudoscalar
transitions is, up to corrections of 
order $\alpha_s(m_H)$ and $\Lambda/m_H\sim \Lambda/E$,~%
\cite{Charles:1998dr,Beneke:2000wa,Bauer:2002aj,Beneke:2003pa,Lange:2003pk,Hill:2004if,Hill:2005ju}
\begin{eqnarray}\label{eq:scetff}
F_+(E) &=& \sqrt{m_H}\big[ \zeta(E) + \left({4E\over m_H} - 1 \right) H(E) \big] \,, \nl
{m_H\over 2E} F_0(E) &=& \sqrt{m_H}\big[ \zeta(E) + H(E) \big] \,,
\end{eqnarray}
where it is more natural to work here in terms of the 
light-meson energy, related to the invariant 
momentum transfer $t$ in (\ref{eq:vectorff}) 
by $t=m_H^2+m_L^2 - 2m_H E$.   
The construction of SCET is more intricate than either HQET or 
CHPT, due to the nonfactorization of large and small momentum modes 
in some processes~\cite{Becher:2003qh,Becher:2003kh}.~%
\footnote{
For explorations along different lines, see \cite{Beneke:2003pa,Manohar:2006nz}. 
} 
Apart from scaling violations related to this phenomenon, the functions $\zeta$ and 
$H$ both have an energy dependence $\zeta\sim H\sim E^{-2}$.

From expression (\ref{eq:scetff}) it is straightforward to see that, independent of 
any model assumptions,~\cite{Hill:2005ju} 
\be
{1\over\beta} = -\left. {d \ln(\zeta + H ) \over d\ln E} \right|_{E=m_H/2} - 1 
+ \order( \alpha_s, \Lambda/m_H ) \,. 
\ee
From the $1/E^2$ asymptotic behavior of $\zeta$ and $H$,  
$1/\beta$ should approach unity as $m_H\to\infty$ at fixed $m_L\lesssim \Lambda_{\rm QCD}$.
Similarly,
\begin{eqnarray}\label{eq:deltascet}
\delta(m_L,m_H) &=& \left. { 2 H \over \zeta + H } \right|_{E=m_H/2} 
+ \order( \alpha_s, \Lambda/m_H ) \,, \nl
 &&\quad \big[ m_H \gg \Lambda \gtrsim m_L \big] \,. 
\end{eqnarray} 
Thus the question of the asymptotic limit, $\lim_{m_H\to\infty}\delta(m_L,m_H)$, 
is the same as the question of which contribution, the 
``hard'' function $H$, or the ``soft'' function $\zeta$, 
dominates in this limit.   

The results summarized by (\ref{eq:deltahqet}), (\ref{eq:deltachiral}) 
and (\ref{eq:deltascet}), can be unified as in Fig.~\ref{fig:contour}, which
displays three equal-$\delta$ contours in the HQET and CHPT regimes. 
The obvious question is: what happens to these contours as we pass 
outside of these regimes? 
We know that $\delta$ cannot depend strongly on $m_H$, since this 
dependence is due solely to scaling violations and power corrections. 
A smooth chiral limit of the form factors also implies that $\delta$ 
cannot depend strongly on $m_L$.
To further constrain these contours, experimental data 
can be used as a quantitative probe
in this regime. 

\section{Implications for phenomenology\label{sec:pheno}}

We focus attention on the prototypical heavy-light process, 
$B\to\pi\ell\nu$.   
The methodology described in Section~\ref{sec:analyticity} 
can be used to extract as much 
information from experimental and lattice data as possible, 
with believable error estimates. 

\subsection{Precision measurements: $V_{ub}$} 

\begin{figure}[h]
\centering
\psfrag{q}[T]{$t$  (${\rm GeV}^2$) }
\psfrag{y}[B]{$10^3|V_{ub}|$}
\includegraphics[width=80mm]{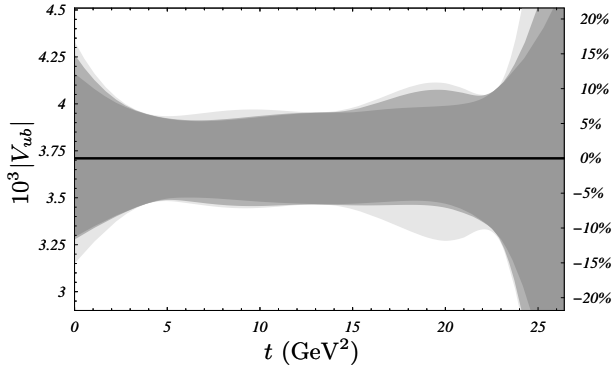}
\caption{ Experimental error on $|V_{ub}|$, with theoretical input of 
the form factor normalization at a given value of $t$, from \cite{Becher:2005bg}.  The plot assumes that the fit yields the central value $|V_{ub}|=3.7\times 10^{-3}$.       
} \label{fig:vub}
\end{figure}

Knowing that the true form factor is given by one of the restricted 
class of curves in (\ref{eq:Fexpand}) allows for maximal use of the 
available experimental~%
\footnote{
For the status of  $B\to\pi\ell\nu$ measurements
see \cite{varvell}
} 
and lattice~%
\footnote{
For recent reviews and references for lattice form factor determinations see
\cite{Wingate:2006ie,mackenzie}. 
}   
data.  
Figure~\ref{fig:vub} shows the minimum error 
obtainable for $|V_{ub}|$ using present 
data~\cite{Athar:2003yg,Abe:2004zm,Aubert:2005cd,Aubert:2005tm}, 
with a form factor determination at a given 
value of $t$~\cite{Becher:2005bg}. 
The dark, medium and light bands correspond to increasing levels of 
conservatism for the size of the coefficients appearing in (\ref{eq:Fexpand}): 
$\sum_k a_k^2 = 0.01, 0.1, 1$, normalized relative to the default unitarity bound.  
At the point $t_0=16\,{\rm GeV}^2$, these values correspond to 
$\sum_k a_k^2/a_0^2 \approx 25, 250, 2500$.  

As Fig.~\ref{fig:vub} illustrates, theory inputs at either very large or
very small $t$ are not as effective as for moderate values, say $t=10-20\,{\rm GeV}^2$, 
which are within the range currently studied with unquenched lattice 
simulations~\cite{Okamoto:2004xg,Shigemitsu:2004ft,Gulez:2006dt}. 
For extreme values of the bounds, e.g., allowing coefficients in the 
expansion (\ref{eq:Fexpand}) to be as large as $a_k/a_0 \sim 50$, the error 
begins to increase, as the lighter bands in the figure show. 

For consistency, the heavy-quark power counting used to establish 
bounds on the form factor shape should be at least as robust and 
conservative as similar estimates used to bound other theoretical errors 
entering a $|V_{ub}|$ determination---e.g.  power corrections, perturbative 
matching corrections, or discretization errors in lattice calculations
of the form factor.
Quantitative investigations such as in Table~\ref{tab:a1} and 
Section~\ref{sec:charm} give us confidence that as far as the bounds are 
concerned, ``order unity'' really means order unity. 

\subsection{Inputs to hadronic $B$ decays} 

Semileptonic decays provide
a robust value for the form factor normalization (times $|V_{ub}|$), 
a key input to factorization analyses of two-body hadronic 
decays~\cite{Bauer:1986bm,Bjorken:1988kk,Beneke:1999br,Beneke:2000ry}. 
From \cite{Becher:2005bg},~%
\footnote{
This may be compared to the  
bound~\cite{Beneke:2003zv} 
$10^3|V_{ub}F_+(0)| < 1.22$ obtained from 
an assumption of form-factor monotonicity and experimental data 
in \cite{Athar:2003yg}, and the value~\cite{Luo:2003hn}
$10^3|V_{ub}F_+(0)|=0.83\pm 0.16$ obtained from
fits of model parameterizations to the same data. 
}
\be
10^3|V_{ub}F_+(0)| = 0.92 \pm 0.11 \pm 0.03 \,.  
\ee

A dominant uncertainty in many factorization predictions is 
the normalization of the hard-scattering contribution to the form factor, 
commonly expressed in terms of  
an (inverse) moment of the $B$ meson wavefunction, $\lambda_B$: 
\be
\delta(m_\pi,m_B) = {6\pi C_F \over N_c} {f_B f_\pi \alpha_s \over m_B \lambda_B F_+(0)} + \dots \,.
\ee
For example, 
the ``default scenario'' inputs of \cite{Beneke:2003zv} 
give $\delta \approx 0.15^{+0.10}_{-0.05}$, while
the ``S2 scenario'' gives a central value $\delta \approx 0.29$.~%
\footnote{
Values are at tree level, 
$\alpha_s(\sqrt{m_b\Lambda}) \approx 0.35$, asymptotic distribution
amplitude for the pion, and the errors shown are from the remaining 
inputs used in
\cite{Beneke:2003zv}. 
}
The semileptonic data can help pin down this number.
From \cite{Becher:2005bg}, using experimental data for $F_+$ from 
\cite{Athar:2003yg,Abe:2004zm,Aubert:2005cd,Aubert:2005tm} determines
\bea\label{eq:deltabpi}
 B\to \pi : \quad\quad\quad && \nl
1+1/\beta - \delta &=& 1.5 \pm 0.6 \pm 0.4 \,, 
\eea
and the lattice value $\beta = 1.2 \pm 0.1$ from \cite{Shigemitsu:2004ft,Okamoto:2004xg}
allows extraction of $\delta$.

If $\delta$ is monotonic as a function of $m_L$, the analysis of Section~\ref{sec:hqet}
shows that 
\be
\delta(m_L, m_B) < 0.35 \pm 0.03 \,, \big[m_L < 1\,{\rm GeV}; {\rm monotonicity}\big] \,. 
\ee 
A significantly larger value of $\delta$, 
e.g. $\delta \sim 1$~\cite{Bauer:2004tj,Bauer:2005kd,Williamson:2006hb}, 
would require a dramatic behavior of the extrapolated curve in 
Fig.~\ref{fig:hqet}.

\section{What's charm got to do with it?\label{sec:charm}}

Charm decays provide a direct probe of the ``interesting'' regime 
pictured in Fig.~\ref{fig:contour}.  They provide an important test 
of lattice measurements for heavy-to-light form factors, and a quantitative
test of the power-counting used to bound the form factor shape in 
other processes such as $B\to\pi$. 

\subsection{Fundamental questions}

\begin{figure}[h]
\centering
\includegraphics[width=80mm]{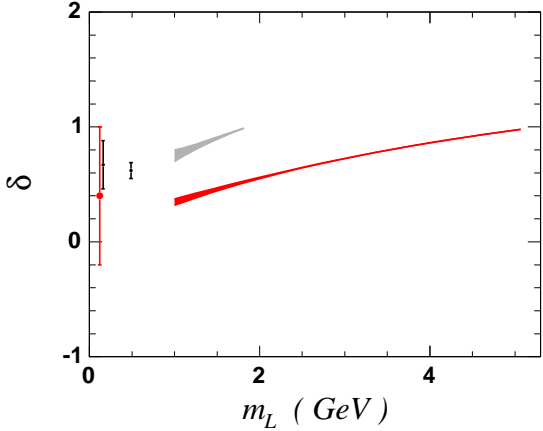}
\caption{
$\delta$ as a function of light-meson mass.  
The upper (black) points are for fixed $m_H=m_D$, and the lower (red) points 
for fixed $m_H=m_B$.    
} \label{fig:deltaBD}
\end{figure}

Depending on whether the soft or hard component of the form factor
dominates in the limit $m_H\to\infty$, the difference in form factor
slopes (\ref{eq:betadef}) tends to $\delta\to 0$ or $\delta\to 2$.  
The latter would require that the curves such as in Fig.~\ref{fig:hqet}
turn upward at small $m_L$, for sufficiently large $m_H$.  
If this happens, and unless a new large scale is dynamically 
generated by QCD, at which a turnaround in 
the curves would occur, some evidence for this
behavior should be evident in the charm system.  
Precision measurements here directly probe the dangerous region 
illustrated in Fig~\ref{fig:contour}. 
For $D\to K$ decays, 
\bea\label{eq:DK}
 D\to K : \quad\quad\quad && \nl 
1+1/\beta-\delta &=& 1.03 \pm 0.09 \pm 0.11 \quad \text{\cite{Huang:2004fr}}  
 \nl
&& 0.94 \pm 0.07 \pm 0.10 \quad \text{\cite{Link:2004dh}} \nl
&& 1.13 \pm 0.10 \pm 0.12 \quad \text{\cite{Abe:2005sh}} \,. 
\eea
Here the first error is experimental using the linear $z$ parameterization 
(just $a_0$, $a_1$) in (\ref{eq:Fexpand}).   The second error is a conservative 
estimate of the residual shape uncertainty, obtained by allowing an extra parameter in
the fit ($a_0$, $a_1$ and $a_2$), with $|a_k/a_0| \lesssim 10$.  
Combining this with the lattice value $\beta = 1.8 \pm 0.1$ yields the $D\to K$
data point in Fig.~\ref{fig:deltaBD}. 

Similarly for $D\to\pi$ decays, 
\bea\label{eq:Dpi}
 D\to \pi: \quad\quad\quad && \nl 
1+1/\beta-\delta &=& 1.3 \pm 0.4 \pm 0.5 \quad \text{\cite{Huang:2004fr}} \nl
&& 0.9 \pm 0.2 \pm 0.3  \quad \text{\cite{Abe:2005sh}} \,, 
\eea
with the errors as above.  Combined with the lattice value 
$\beta = 1.65 \pm 0.10$ yields the $D\to\pi$ 
data point in Fig.~\ref{fig:chiral} and 
Fig.~\ref{fig:deltaBD}.

Our current combined knowledge of form factor shape 
from $B$ and $D$ decays is illustrated in 
Fig.~\ref{fig:deltaBD}.   So far the data do not indicate 
surprises in either of the curves when extrapolated 
into the region $m_L \lesssim 1\,{\rm GeV}$.  It will be interesting to probe
this region with more precision when further data becomes available.

\subsection{Lattice, experiment and parameterizations} 

Charm decays provide important tests of nonperturbative methods 
used to evaluate hadronic matrix elements.  
When comparing the results of lattice QCD with experiment, it should 
be kept in mind that the kinematic regions that are studied with 
best precision are different for the lattice (large $t$) and 
experiment (small $t$).  Also, the manner in which chiral extrapolations
are performed to reach physical light-quark masses imply that it is 
difficult to present lattice results in terms of uncorrelated values of
the form factor at different $t$ values.  In practice, the results are 
generally presented in terms of a  parameterized curve; to make a definitive 
comparison to experiment, it is essential that the chosen parameterization 
doesn't introduce a bias.
The ideas described in Section~\ref{sec:analyticity} allow a systematic
approach to this problem~\cite{mackenzie}. 
The remainder of this subsection points out pitfalls that can occur 
with some of the simplified parameterizations in common use. 

The starting point for many parameterizations is
a more pedestrian but rigorous approach to analyticity, which implies 
the dispersion relation: 
\begin{equation}\label{eq:dispersion}
F_+(t) = {F_+(0)\over 1-\alpha}{1\over 1- t/m_{H^*}^2} +  
{1\over \pi}\int_{t_+}^\infty {dt^\prime} { {\rm Im}F(t^\prime) \over t^\prime -t} \,,  
\end{equation}
where a distinct $m_{H^*}$ pole appears below threshold for heavy-to-light decays
such as $B\to\pi$ and $D\to K$ (and almost for $D\to\pi$).  
The first interesting
test is to see whether just the $m_{H^*}$ pole can describe the data, 
\begin{equation}\label{eq:simple} 
\begin{array}{ll}
{\rm vector\,\, dominance }: \\ 
& \displaystyle{ F_+(t) = {F_+(0)\over 1 - t/m_{H^*}^2 } } \,.  \quad\quad\quad\quad\quad\hspace{5cm}\phantom{a} \\  
\end{array} 
\end{equation}
In fact this ``vector dominance'' model can be explicitly 
ruled out by the
data~\cite{Link:2004dh,Huang:2004fr,Abe:2005sh,Abe:2005sh,Athar:2003yg,Aubert:2005cd,Abe:2004zm},
so that inclusion of the continuum contribution in (\ref{eq:dispersion}) 
is essential.  

From a dynamical point of view,  
in order to obtain the $1/E^2$ dependence appearing in heavy-to-light 
form factors, as in (\ref{eq:scetff}), 
it is necessary that the continuum integral in (\ref{eq:dispersion}) 
play a significant role.  This can be treated in a model independent way by 
breaking up the integral into a sum of effective poles, and using power-counting
estimates to establish reasonable bounds on the coefficients of these effective 
poles~\cite{Hill:2005ju,Becher:2005bg}.
In the first approximation, the continuum integral is represented by a single 
effective pole, and two parameters are necessary to describe its location and
strength relative to the $H^*$ pole.  Since this is one more parameter than is 
easily measured from the data, various suggestions have been made for eliminating 
one of these parameters.   

The ``single pole model'', where both the $H^*$ pole and the continuum integral 
are represented by a single pole that is allowed to float,
\begin{equation}\label{eq:single}
\begin{array}{ll}
{\rm single\,\, pole }: \\ 
& \displaystyle{ F_+(t) = {F_+(0)\over 1 - t/m_{\rm pole}^2 } }\,, \quad\quad\quad\quad\quad\quad\quad \\
\end{array} 
\end{equation}
is also ruled out by the data, although in a slightly less direct way. 
Although the curve can be made to fit, the pole position is forced to take an 
unphysical value, significantly below both the $H^*$ pole and the continuum.
The ``modified pole model'', 
\begin{equation}\label{eq:modified}
\begin{array}{ll}
{\rm modified\,\,pole }:  \\
& \hspace{-10mm} \displaystyle{ F_+(t) = {F_+(0)\over \left( 1 - t/m_{H^*}^2\right)\left(1-\alpha_{\rm pole} t/m_{H^*}^2\right) } } \,,
\end{array} 
\end{equation} 
holds a similar status. 
For $D\to K$ and $D\to \pi $, (\ref{eq:DK}) and (\ref{eq:Dpi}) clearly show that 
$1+1/\beta -\delta \approx 2$ is not valid for charm decays, 
as necessary for the motivation for the 
simplification (\ref{eq:modified}) proposed in \cite{Becirevic:1999kt}.
Although the form (\ref{eq:modified}) can be made to fit 
the data, there is no obvious physical interpretation for the resulting fit 
parameter; in particular $\alpha_{\rm pole}$ obtained 
in this way has no direct relation to the physical $\alpha$ defined in 
(\ref{eq:dispersion}).   

It should be kept in mind that unless there is a physical meaning that can 
be given to the parameter being studied, there is no guarantee that different
experimental or lattice determinations will converge to any one value for this parameter.  
It is therefore unclear what to make of discrepancies appearing when different
datasets are forced to fit models such as (\ref{eq:single}) 
or (\ref{eq:modified})~\cite{Wiss:2006ih,poling}.  
The situation is especially dangerous for comparing lattice and experiment, since
the range of $t$ that is emphasized is different in the two cases.  
These pitfalls are easily avoided by working with a general  
parameterization such as (\ref{eq:Fexpand}) that is guaranteed to contain the 
true form factor, and by comparing physical quantities, such as in (\ref{eq:DK}), 
(\ref{eq:Dpi}).

\subsection{Testing convergence} 

The general parameterization (\ref{eq:Fexpand}) 
provides a systematic procedure for estimating
how many terms should be resolved by data at a given level of precision. 
Rigorous bounds are placed on the coefficients by using crossing symmetry 
to analyze the production form factor, either via unitarity arguments, or 
through power counting of contributions from different momentum regions. 
Since the latter estimates yield constraints that are so much more 
powerful than can be safely estimated by unitarity, it is important
to check wherever possible that large ``order unity'' numbers don't 
appear.
As illustrated by Table~\ref{tab:a1}, the available semileptonic data 
reveal no surprises. 

\section{Decays of pseudoscalar to vector mesons\label{sec:pv}}

\begin{figure}[h]
\centering
\includegraphics[width=75mm]{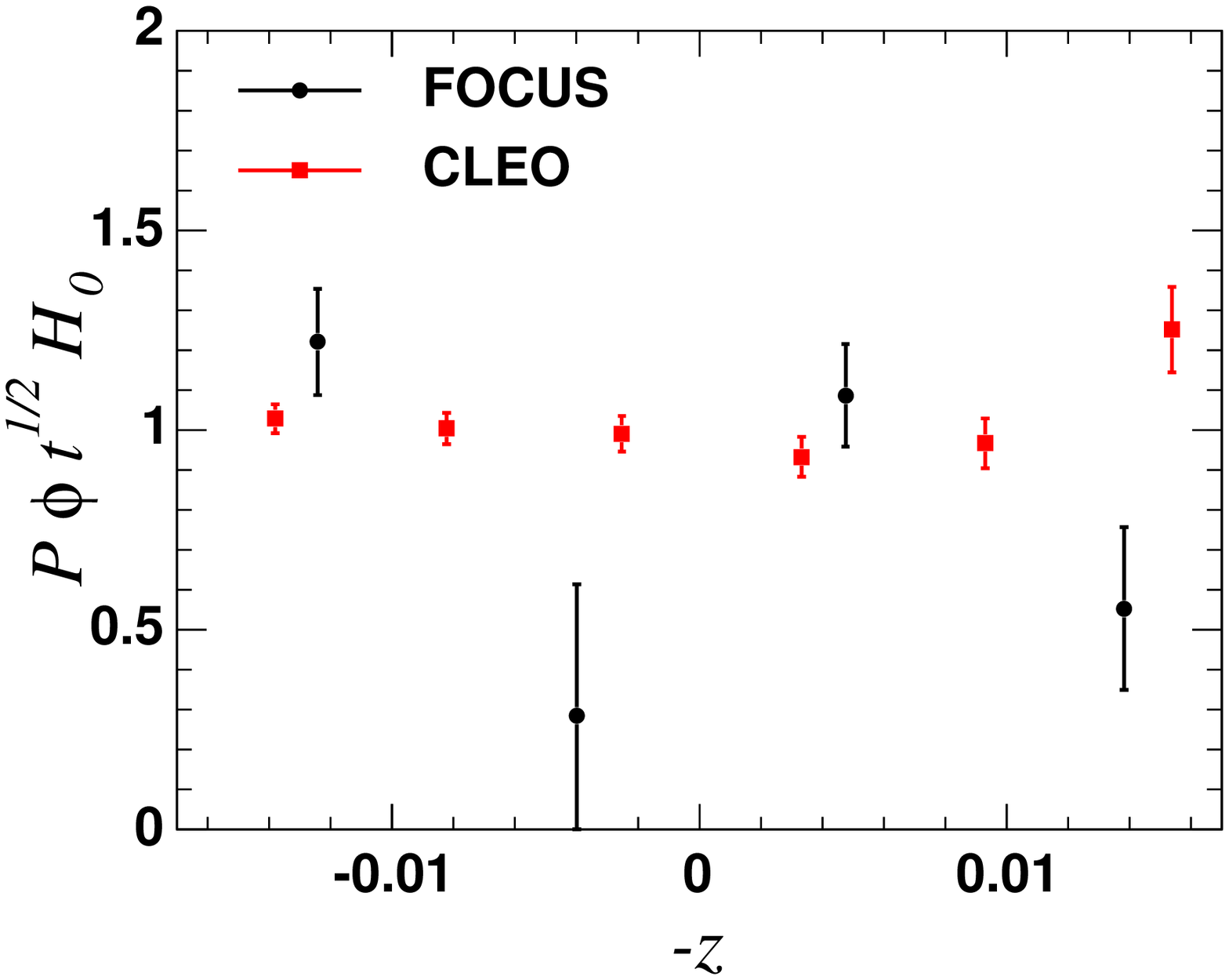}
\includegraphics[width=75mm]{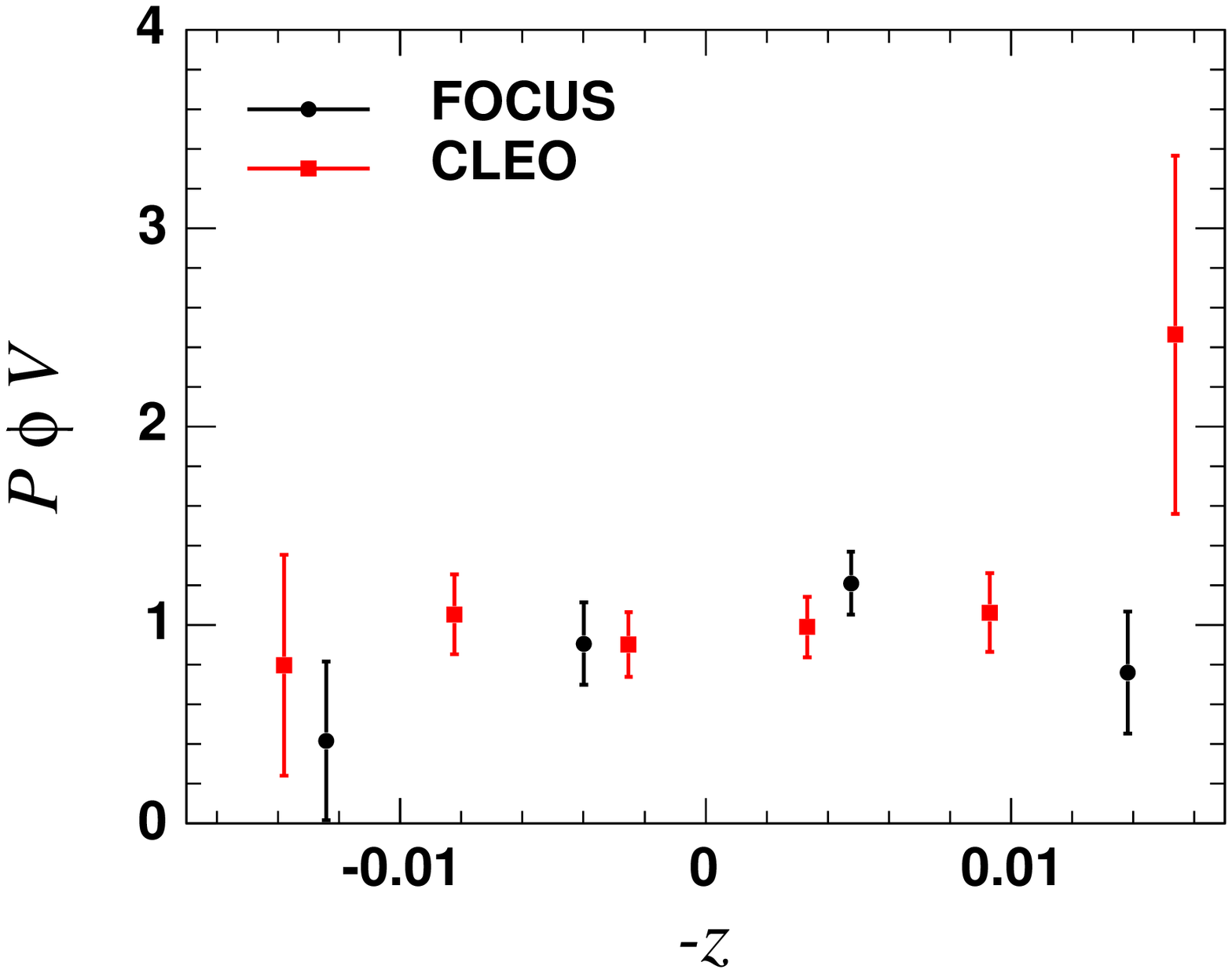}
\includegraphics[width=75mm]{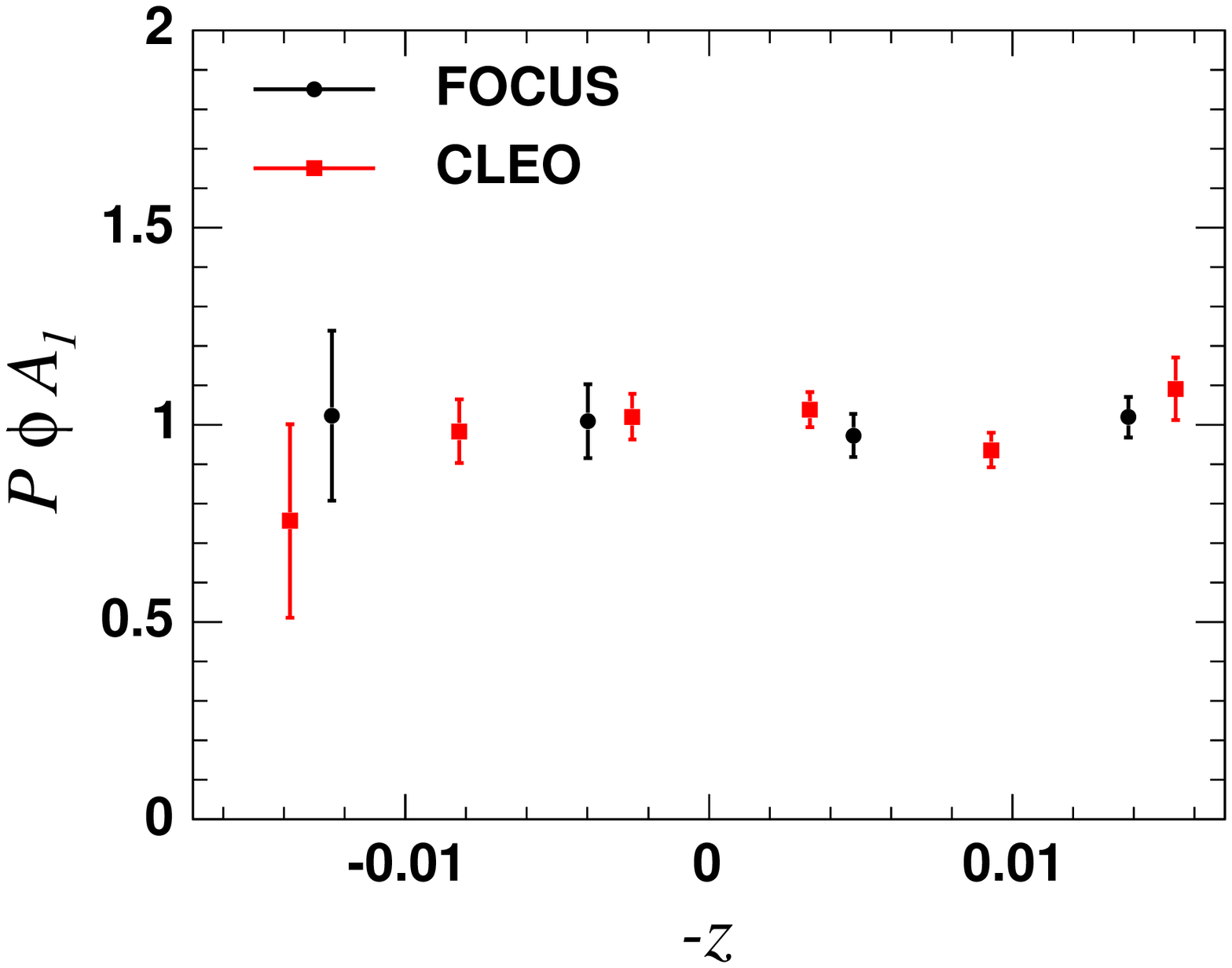}
\caption{Experimental data for helicity amplitudes 
in $D\to K^*\ell\nu$, from \cite{Link:2005dp,Shepherd:2006tw}.  
Units on the vertical axis are arbitrary 
(approximately normalized to unity at $t=0$).  
 } \label{fig:dkstar}
\end{figure}

Pseudoscalar-pseudoscalar transitions hold a priveleged position, 
from a first-principles simplicity point of view, from the lattice point
of view, and from the experimental point of view.
Pseudoscalar-vector transitions, while accompanied by new complications, 
are however important backgrounds to the pseudosclar mode, provide alternative 
extractions of CKM parameters, and yield important constraints on 
radiative and hadronic transitions. 
This section briefly outlines the implementation of the ideas in 
Section~\ref{sec:analyticity} to the pseudoscalar-vector case.~%
\footnote{
There have been numerous studies aimed at reproducing the $P\to V$ 
data by means of symmetry arguments or proposed generating 
resonance structures~\cite{Fajfer:2005ug,Fajfer:2006uy,Ebert:2001pc}.  
The focus of the present talk is on the extraction of physical quantities without 
simplifying or model assumptions. 
}

The most obvious complication is the multiple invariant form factors 
that accompany the vector particle.   
Less obvious complications involve modifications to the analytic structure
of the form factors due to the unstable nature of vector mesons, and the 
possible existence of anomalous thresholds.  When these 
occur, they will encroach on the gap between the semileptonic region and 
singularities.  
The effects of such anomalous thresholds are not expected to be large, and 
can be investigated on a case-by-case basis.~%
\footnote{
For the case of $B\to D^*\ell\nu$, see e.g. \cite{Boyd:1995sq,Caprini:1995wq}.  
}
A more complete discussion is beyond the scope of this talk
and we ignore such complications here. 

\begin{table}[t]
\begin{center}
\caption{Maximum $|z(t,t_0)|$ throughout semileptonic range 
with symmetrizing choice $t_0=t_+(1-\sqrt{1-t_-/t_+})$.}
\begin{tabular}{lc}
\textbf{\quad Process \quad} & \textbf{$|z|_{\rm max}$}
\\
\hline 
$D\to K^*$ & 0.017 \\
$D\to \rho$ & 0.024 \\
$B\to D^*$ & 0.028 \\
$B\to \rho$ & 0.10 \\
\end{tabular}
\label{tab:zvector}
\end{center}
\end{table}

In heavy-to-heavy decays it is possible to relate form factors by heavy-quark
symmetry, making the pseudoscalar-vector analysis not significantly different, 
from the point of view of the number of independent invariant form factors,
from the pseudoscalar-pseudoscalar case.   
For heavy-to-light decays, this simplification is not possible; however, a new
symmetry emerges due to the large energy of the vector meson, and the resulting
suppression of the helicity-flip 
amplitude~\cite{Burdman:2000ku,Beneke:2000wa,Hill:2004if}.   
It is again important to take advantage of as much first-principles knowledge 
as possible.  By the same variable transformation (\ref{eq:zdef}), 
the invariant form factors may all be written in terms of a convergent expansion in 
a small parameter governed by the degree of isolation of the semileptonic 
region.   Table~\ref{tab:zvector} shows the maximum 
size of the parameter when $t_0=t_+(1-\sqrt{1-t_-/t_+})$, for processes involving
ground state heavy-light pseudoscalar mesons decaying into the lowest-lying 
vector mesons. 

The decay rate can be decomposed in terms of helicity amplitudes~\cite{Korner:1989qb},~%
\footnote{
Form factor conventions are as in \cite{Hill:2004if,Hill:2004rx}. 
} 
\bea
{\sqrt{t} H_0\over m_H^2} 
&=& { (1+\hat{m_L})^2(\hat{E}_L-\hat{m_L}^2)A_1 - 2\hat{k}_L^2 A_2 \over \hat{m}_L (1+\hat{m}_L) } \,, \nl
{H_\pm \over m_H} &=& 
{ (1+\hat{m}_L)^2 A_1 \mp 2 \hat{k}_L V \over 1 + \hat{m}_L } \,,
\eea
where hatted variables are in units of $m_H$, and $E_L$ and $k_L$ denote the
energy and momentum of the light vector meson in the rest frame of the 
heavy pseudoscalar meson.  
The amplitudes $\sqrt{t} H_0$, $V$ and $A_1$ 
 correspond to
cross-channel production of states with quantum numbers 
$J^P = 1^+$, $1^-$ and $1^+$, respectively. 
On the premise of analyticity, these functions can be expanded as
in (\ref{eq:Fexpand}).   For the choice of $\phi$, we use the form 
motivated by unitarity (although the evaluation of the unitarity bound 
itself is not directly relevant). 
Using the results of \cite{Caprini:1997mu}, at $Q=0$,  
\bea
\phi_{\sqrt{t} H_0}(t,t_0) &\propto& 
{\sqrt{t_+ - t}\over 
(t_+ - t_0)^{1/4} }
\left( z(t,0)\over -t \right)^{5/2} 
\left( z(t,t_0) \over t_0-t \right)^{-1/2} \nl
&& \times 
\left( z(t,t_-) \over t_--t \right)^{-1/4}   \,,\nl
\phi_{V}(t,t_0) &\propto& 
{ t_+ - t \over (t_+ - t_0)^{1/4}}
\left( z(t,0)\over -t \right)^{5/2} 
\left( z(t,t_0) \over t_0-t \right)^{-1/2} \nl
&&\times
\left( z(t,t_-)\over t_- - t \right)^{-3/4} \,, \nl
\phi_{A_1}(t,t_0) &\propto& 
{ \sqrt{t_+ - t}\over 
(t_+ - t_0)^{1/4} } 
\left( z(t,0)\over -t \right)^{2} 
\left( z(t,t_0) \over t_0 - t \right)^{-1/2} \nl
&&\times 
\left( z(t,t_-) \over t_--t \right)^{-1/4} \,. 
\eea
For the function $P$, we use for the resonances 
lying below threshold:~\cite{DiPierro:2001uu,Eichten:2005sf} 
$J^P=1^-: m = 2.11, 2.63\,{\rm GeV}$; 
$J^P=1^+: m = 2.46, 2.56\,{\rm GeV}$.~%
\footnote{
Fortunately from the present point of view (unfortunate for
probing these states), the fits are not very sensitive to the location
or even existence of these additional poles. 
}  

Fig.~\ref{fig:dkstar} shows the resulting invariant form factors extracted 
from the nonparametric analyses in \cite{Link:2005dp,Shepherd:2006tw}, 
after extracting the kinematic function $P\phi$, and transforming to 
the variable $z$.    
The small size of $z$ implies that these functions
should deviate from a straight horizontal line by only a few percent 
over the entire range.  
With this in mind, it is straightforward to extract useful
information in a systematic way.  For instance, we can test the 
$\order(\Lambda_{\rm QCD}/m_c)$ suppression of helicity amplitudes at $t=0$:
[equivalently, measure the form factor ratio $V(0)/A_1(0)$] 
\bea
{H_+\over H_-} &=& 0.27 \pm 0.06 \quad \text{\cite{Link:2005dp} } \nl 
&& 0.37 \pm 0.04 \quad \text{\cite{Shepherd:2006tw} } \,, 
\eea
up to corrections of $\order(z)$. 

\section{Summary} 

There is interesting information lurking in the semileptonic data, 
with applications to precision phenomenology, and for 
exploring a poorly-understood limit of QCD. 
As a practical matter, a simple and well-known 
variable transformation provides a powerful tool for analyzing
and making full use of the semileptonic data.  
The effectiveness of this transformation has been 
obscured by a reliance on unitarity bounds, which were used as
safe estimates in the absence of experimental data. 
Rigorous arguments for a stronger convergence
of the expansion (\ref{eq:Fexpand}) are now available, 
and can be further tested and refined using experimental 
data from many semileptonic decay modes. 

\bigskip 
\begin{acknowledgments}
It is a pleasure to thank the organizers for an enjoyable conference. 
Thanks also to T.~Becher for discussions, and collaboration on 
Ref.~\cite{Becher:2005bg}, on which most of Section~\ref{sec:pheno} of this
talk is based, 
to E.~Eichten for clarifying the situation with 
$D_s$ states in \cite{DiPierro:2001uu,Eichten:2005sf}, 
and A.~Kronfeld and P.~Mackenzie  for 
discussions relating to the lattice. 
Fermilab is operated by Universities Research Association Inc. 
under contract with the U.S. Department of Energy. 
Research supported by Grant DE-AC02-76CH03000. 

\end{acknowledgments}

\end{document}